# Influence of local, residual structure on the scaling behavior and dimensions of unfolded proteins

Zhisong Wang*[1,3], Kevin W. Plaxco[2], and Dmitrii E. Makarov[3]

[1]Institute for of Modern Physics and Applied Ion Beam Physics Laboratory, Fudan University, Shanghai 200433, China

[2]Department of Chemistry & Biochemistry, University of California, Santa Barbara, CA 93106

[3]Department of Chemistry & Biochemistry and Institute for Theoretical Chemistry, University of Texas at Austin, Austin, TX 78712

*Correspondence should be addressed to: Z. Wang

Institute of Modern Physics, Fudan University

Shanghai 200433, China

Phone: ++86 21 55664592
Fax: ++86 21 65642787
E-mail: wangzs@fudan.edu.cn

**Abstract**

Whereas recent spectroscopic studies of chemically denatured proteins hint at significant non-random residual structure, the results of extensive small angle x-ray scattering studies suggest random coil behavior, calling for a coherent understanding of these seemingly contradicting observations. Here we report the results of a Monte Carlo study of the effects of two types of local structures, $\alpha$ helix and Polyproline II (PPII) helix, on the dimensions of random coil poly-alanine chains viewed as a model of highly denatured proteins. We find that whereas Flory's power law scaling, long regarded as a signature of random coil behavior, holds for chains containing up to 90% $\alpha$ or PPII helix, the absolute magnitude of the chain dimensions is sensitive to helix content. As residual $\alpha$ helix content increases, the chain contracts until it reaches a minimum radius at ~70% helix, after which the chain dimensions expand rapidly. With an $\alpha$ helix content of ~20%, corresponding to the Ramachandran probability of being in the helical basin, experimentally observed radii of gyration are recovered. Experimental radii are similarly recovered at an $\alpha$ helix content of ~87%, providing an explanation for the previously puzzling experimental finding that the dimensions of the highly helical methanol-induced unfolded state are experimentally indistinguishable from those of the helix-poor urea-unfolded state. In contrast, the radius of gyration increases monotonically with increasing PPII content, and is always more expanded than the dimensions observed experimentally. These results suggest that PPII is unlikely the sole, dominant preferred conformation for unfolded proteins.

**INTRODUCTION**

Growing interest in the unfolded state is reflected in numerous recent spectroscopic, small angle scattering and simulations-based studies of the properties of highly denatured proteins. Emerging from many of the spectroscopic studies is the remarkable observation of native-like long-range order[1-3] as well as evidence of persistent local structure and hydrophobic clustering under even highly denaturing conditions[4, 5]. Similarly, several recent spectroscopic and simulation-based studies suggest that poly-proline II (PPII) structure is a preferred backbone conformation for short, unfolded peptides and may be a significant contributor to fully denatured ensembles[6-9]. These assertions challenge the long-standing perception that the unfolded state of proteins is effectively a random coil state lacking any specific, well-populated structures[10]. In contrast to these reports of residual denatured state structure in even the most highly denatured proteins, systematic experimental studies based on small angle x-ray scattering lend strong support to the random coil picture[11]. It has been found[11, 12] that the unfolded states generated by a range of denaturing conditions and for a wide variety of arbitrarily chosen proteins, exhibit robust, systematic features that are completely consistent with theoretical models of the random coil state. Perhaps most notably among these, recent studies[13] have shown that the radius of gyration ($R_g$) of denatured proteins varies as a function of the length of the unfolded protein (residue number, $N$) via the $R_g \sim N^{3/5}$ power law predicted by Flory for random-coil polymers in a good solvent[14]. A partial reconciliation of these seemingly contradicting findings is provided by Fitzkee and Rose[15], who have shown by simulation that even ensembles that contain significant, native-like local structure can recover experimentally-observed, denatured-state radii. Here we explore the interplay between sequence-local structure and the global dimensions of the unfolded state in more detail.

In the present study, we focus on the influence of residual α or PPII helices on the random coil behavior of unfolded proteins. Our interest in α-helices stems from the fact that they have been found to exist abundantly in the unfolded state induced by organic co-solvents, such as methanol, presumably because the organic solvents tend to reduce polypeptide-water interactions and thus favor intra-molecular hydrogen bonding[16]. In contrast, urea and GuHCl are thought to disrupt intra-chain hydrogen bonds[17], leading to an unfolded state of much lower α helix population[4, 5]. Considering the significant difference in helix abundance in the urea/GuHCl-unfolded state relative to the alcohol-unfolded state, the experimental finding that the dimensions of the two unfolded states are virtually indistinguishable [11] is rather puzzling. Our interest in PPII, which is a three-fold left-handed helix that was first recognized in proline-rich peptides, stems from recent claims that PPII is a preferred configuration for unfolded polypeptides. Evidence in favor of this assertion includes the observation that the circular dichroism (CD) spectrum of a pure PPII helix is quite similar to the spectra of denatured proteins, a fact that led Tiffany and Krimm to their much disputed, now thirty-year-old hypothesis that PPII segments are populated in the denatured state[18]. More recently NMR experiments have identified PPII as the dominant conformation for a short alanine peptide[9], renewing the interest in PPII structure [6, 8, 19]. To date, however, significant amounts of PPII helix have only been observed for short peptides[6, 7, 9], in which hydrogen bonding is insufficient to stabilize the α helix and the peptide backbone tends to expose itself maximally to the surrounding solvent. The extent to which the PPII helix occurs in chemically or thermally denatured proteins of large *N* has not been established. Using a Monte Carlo algorithm to generate ensembles of polypeptides with specified residual helix contents, we explore here the influence of residual α and PPII helices on the scaling behavior and chain dimensions of denatured proteins.

**METHODS**

In order to address the extent to which residual α or PPII structure affects the random coil dimensions of unfolded proteins, we adopted a Monte Carlo (MC) simulation method based on an all-atom representation of polyalanine chains[20-22]. A simplified model treating helices as rods and random coil segments as Gaussian chains is used to gain additional insight into the MC results.

**Monte Carlo Simulation**

Proteins can be denatured via the addition of chaotropes such as urea or guanidine hydrochloride, or by changing thermodynamic conditions such as temperature or pressure. The different denaturing methods lead to different intra-chain and chain-solvent interactions, which are poorly understood and difficult to quantify[4, 23-25]. Here we simulate the unfolded state by "turning off" all attractive interactions within the chain and using a simple all-atom, polyalanine model with a hard-shell repulsion potential. To generate putative conformations of such chains, we use a Monte Carlo procedure in which $\{\Phi,\Psi\}$ values are randomly selected for each residue from a Ramachandran map. Conformations violating the steric constraints are discarded. The allowed $\{\Phi,\Psi\}$ values in the Ramachandran map are those satisfying the hard shell repulsion steric constraints in the alanine dipeptide (Fig.1). Such Ramachandran plots are a historical standard in protein simulations[10, 26, 27]. Values for hard-sphere contact distances in this work are presented in Table 1.

The likelihood of a randomly generated conformation to violate steric constraints increases with the chain length. For this reason, it is a challenging computational task to generate a statistically sufficient number of conformations for long polypeptides such as $N = 550$ as studied in denaturing experiments. We have employed the highly efficient pivot algorithm [28], which allowed us to easily accumulate more than $1 \times 10^6$ conformations for chains as long as $N = 550$ without biasing the sampling statistics.

We measure the α or PPII helical content in the simulated ensembles relative to a reference, "maximally unfolded", state that is generated by selecting $\{\Phi,\Psi\}$ pairs from the Ramachandran map according to a uniform probability distribution. To generate ensembles with an arbitrary helix content, we bias the $\{\Phi,\Psi\}$ preferences of this unstructured chain ensemble towards α or PPII. This is done by dividing the Ramachandran map into two regions, one corresponding to the helical conformation of interest and the other covering all remaining dihedral angle pairs. The ideal α helix is given by the dihedral angle $\{\Phi = -57.8°, \Psi = -47°\}$, and the ideal PPII helix by $\{\Phi = -75°, \Psi = 145°\}$. The two helical regions of the Ramachandran map are defined as all $\{\Phi,\Psi\}$ within ± 1.8° of the ideal α helix and PPII values.

In the pivot algorithm[28], an initial, equilibrated chain conformation is generated for a fixed chain length $N$. Then a peptide unit is randomly selected, and given new $\{\Phi,\Psi\}$ values, resulting in a different conformation that is accepted (rejected) provided it lacks (contains) steric overlaps. To generate statistically independent conformations, special cares must be taken to discard a certain number of conformations at the initialization stage and during the runs. A detailed description of the pivot algorithm can be found in ref. 28. To bias this procedure towards generating helix-containing structures, the initial conformation is built by adding peptide units one by one, with the $\{\Phi,\Psi\}$ pairs being assigned the helical values according to an input helical probability, $P_h'$. Because of steric restrictions, the actual helical content of the resultant conformation, $P_h$, may differ from the input value of helical probability. Here we define the helical content of an individual conformation as the fraction of the {Φ,Ψ} pairs falling within the helical region of the Ramachandran plot. The subsequent pivot moves are also biased using the desired helical content. The chosen peptide unit is assigned the helical {Φ,Ψ} values according to a probability of

$$P = \begin{cases} 1 & P_h < P_h' - \Delta P_h \\ \frac{1}{2}\left(1 - \frac{P_h - P_h'}{\Delta P_h}\right) & P_h' - \Delta P_h < P_h < P_h' + \Delta P_h \\ 0 & P_h > P_h' + \Delta P_h \end{cases},$$

where $P_h$ is the actual helical content of the parent conformation in the pivot series, and $\Delta P_h$ is an allowed deviation from the desired helical content $P_h'$. $P_h$ is recorded for each individual conformation and the average $P_h$ over the whole ensemble, $<P_h>$, is used in the final analysis. The biased procedure guarantees that the resulting ensemble has a mean helical content within $P_h' \pm \Delta P_h$. As a check of consistency of the procedure, we found that the value of $<P_h>$ always converges to the input value $P_h'$ after a sufficient number of conformations are sampled. We used $\Delta P_h = 5\%$ throughout this paper. The properties of the generated ensemble are rather insensitive to the precise value of $\Delta P_h$. In particular, the mean value of the radius of gyration $R_g$ remained virtually unchanged when a larger value, $\Delta P_h = 20\%$, was used instead.

The above definition of helical content essentially quantifies how frequent the backbone dihedral angles adopt the values for α or PPII conformation. This definition of helical content is not unique; it is an ensemble average that is defined to within $\pm \Delta P_h$. Further, our definition adopts an arbitrarily chosen boundary between the helical and non-helical conformations in the Ramachandran map. The size of the regions we have defined as α or PPII helix is rather restrictive: the 'non-helical' region so defined contains some dihedral angles that may actually lead to conformations quite similar to α or PPII helices. By this treatment the α and PPII helices in the computer generated ensembles will be reflected by the values of ⟨$P_h$⟩ with an error. However, this error is within the overall error bar of $\Delta P_h$ = 5% imposed by the sampling method. To make this clear, let us take for example the ensemble generated with an initially imposed value of $P_h'$ = 20% for α helical content. We note that the value of helical content actually used in the final analysis is not the imposed value ($P_h'$) but the value of ⟨$P_h$⟩ that was obtained by directly counting over chain conformations in the generated ensemble. Our calculations show that the counted value is generally within 3% of the initially assigned value. Suppose the counted value is found to be 22% for the ensemble. This means that 22% of the backbone dihedral angles adopt the values from the narrow region we defined for ideal alpha or PPII conformation. The other 78% of the backbone dihedral angles were taken from the entire Ramachandran map according to a uniform probability distribution. Thus the 78% dihedral angles may adopt values from the broadest possible {Φ,Ψ} basin for alpha conformation. But the even sampling of Ramachandran map guarantees that no more than 5% of the 78% dihedral angles adopt alpha conformation, even though the alpha basin based on known protein structures is considered. Therefore, the 78% backbone dihedral angles introduce an error of less than 4% for the helical content. This level of uncertainty is within the overall error bar of our sampling method. So use of a broader basin for the 22% dihedral angles does not lead to more alpha helices in the generated ensemble, rather the same amount of alpha segments take slightly different 3-dimentional structures. Therefore, the current estimation of helical contents and the error bars are reliable in supporting the conclusions of the present study. Essentially, our method provides an unambiguous measure of how much *more* α or PPII helices are introduced relative to the unstructured chain ensemble, thereby enabling us to study the effect of this additional residual structure on the behavior of the denatured ensemble.

**Broken Rod-like Chain (BRC) Model**

The effect of helical content on the average chain dimensions can be understood qualitatively by using the broken rod-like chain model[29, 30]. In this model, one ignores excluded volume and approximates helical segments as rigid rods, which are interrupted by random coil segments. Consider a chain that consists of $N \gg 1$ links. The mean square end-to-end distance of such a chain is given by

$$\langle R^2 \rangle \approx b^2 \langle n_c \rangle + l_h^2 \langle N_h \rangle \langle n_h^2 \rangle.$$

Here $\langle n_c \rangle$ is the average *total* number of residues in the random coil state, $\langle N_h \rangle$ is the average number of helical segments in the chain, $l_h$ is the helical rise per residue (1.5 Å for $\alpha$ and 3.1 Å for PPII helix), and $b$ is a quantity characterizing the random coil conformations: $<R^2> = Nb^2$ where $R$ is the end-to-end distance of the random coil chain. The quantity $\langle n_h^2 \rangle$ is the average squared number of monomers $n_h$ in a single helical segment.

In general, $\langle N_h \rangle$ and $\langle n_h^2 \rangle$ depend not only on the average helical content but also on the statistics of helical segments in the chain. In the Zimm-Bragg model[31], the probability p($n_h$) of having $n_h$ adjacent helical segments depends on two parameters, the helix initiation (or "cooperativity") parameter $\sigma$ and the helix propagation parameter $s$. The distribution of the length of helical segments is Poisson:

$$p(n_h) \propto (s/\lambda)^{n_h},$$

where $\lambda = (1 + s + \sqrt{(s-1)^2 + 4s\sigma})/2$, so that

$$\langle n_h^2 \rangle = \sum n_h^2 p(n_h) = \frac{(1+s/\lambda)}{(1-s/\lambda)^2}.$$

The average number of helical segments can be estimated as

$$\langle N_h \rangle = N\theta_h / \langle n_h \rangle,$$

where $\theta_h = \frac{1}{2} + \frac{s-1}{2\sqrt{(s-1)^2 + 4s\sigma}}$ is the average helical content and

$$\langle n_h \rangle = \sum n_h p(n_h) = \frac{1}{1-s/\lambda}$$

is the average number of residues in a helical segment.

If one further assumes that there is no cooperativity (i.e., $\sigma = 1$) then $\lambda = s + 1$, $\theta_h = s/(s+1) = P_h$, $\langle n_c \rangle = N(1-P_h)$. In this limit, each residue assumes the helical conformation with the probability $P_h$ and a coil conformation with a probability $P_c = 1 - P_h$, and all residues are statistically independent. The above equations then give:

$$\langle R^2 \rangle \approx N\left[b^2(1-P_h) + l_h^2 P_h \frac{1+P_h}{1-P_h}\right] \equiv Nb_{eff}^2. \tag{1}$$

This equation suggests that long chains effectively behave as random coils, with a modified value of $b$ that depends on $P_h$. The ratio of $R_g$ for a finite value of helical propensity over that for the pure coil state can then be approximated as:

$$\frac{R_g(P_h)}{R_g(0)} = \left[(1-P_h) + \frac{l_h^2}{b^2}\frac{P_h(P_h+1)}{1-P_h}\right]^{1/2}. \tag{2}$$

While replacing helices by rods and ignoring excluded volume effects are potentially rather drastic approximations, this model captures key features of the dependence $R_g(P_h)$. It further allows one to assess the effects of cooperativity, by changing the parameter $\sigma$, although those are beyond the scope of the present paper.

**RESULTS AND DISCUSSION**

**The Random-coil Statistics May Coexist with A Significant Presence of Residual Sequence-local Structures**

Studies of denatured globular proteins showed that the dimensions of chemically and thermally unfolded proteins exhibit a saturation behavior: the radii of unfolded proteins do not further increase upon increasing temperature or guanidine or urea concentration [11]. Furthermore, all three denaturing methods ultimately lead to virtually indistinguishable chain dimensions[11]. This is true for the very large majority of unfolded proteins characterized to date that span a broad range of chain length $N$ (up to $N = 416$) and amino acid sequences. Recent experimental analysis[13] of almost thirty guanidine- or urea-denatured proteins indicates that $R_g$ scales with chain length according to a power law

$$R_g = R_0 N^v, \qquad (3)$$

with $v = 0.598 \pm 0.029$, extremely close to the 3/5 power predicted by the Flory theory for excluded-volume random-coil polymers in a good solvent. The robustness of the 3/5 power is also shown by a recent theoretical study by Goldenberg[32] for four proteins with $N$ ranging from 26 to 268: he finds $v = 0.58 \pm 0.02$ for an ensemble of Monte Carlo generated unfolded conformations. Including longer chains, $N \leq 600$, we find in the present study $v = 0.605 \pm 0.005$ for the random coil ensemble ($P_h = 0$) of poly-alanine chains (Fig. 2B). Increasing $\alpha$ or PPII helix content up to $P_h = 90\%$, the power law scaling and the 3/5 exponent remain largely intact (Fig. 2A, B). In Fig. 2B we plot the fitted value of $v$ as a function of the helical content and see that it deviates by less than 5% from Flory's value for $P_h = 0 - 90\%$. These findings are of course consistent with the predictions of scaling theory[33]. By increasing the helical content one increases the average length of ordered, helical segments of the chain. This effectively increases the persistence length of the chain (*cf*. Eq. 1). However because the helical segments in the chain are separated by random, non-helical residues, as long as the overall chain contour length is much longer than the persistence length one should recover the standard scaling behavior. Once the helical content becomes so high that the length of ordered segments is comparable with the chain length $N$ then one would see deviations from the Flory scaling law. Specifically, the effective value of $\nu$ should become larger going towards the rigid rod scaling ($\nu=1$). Indeed, we see this trend in Fig. 2B. The above argument suggests that the Flory scaling should be immune to any residual *sequence-local* structures dispersed throughout the polypeptide chain. Thus, the Flory scaling, long regarded as a benchmark signature of random-coil behavior, can naturally coexist with non-trivial populations of sequence-local structures[15]. As it has turned out in denaturing experiments, while the unfolded states of many proteins host a spectroscopically non-trivial amount of structural order, this order does not significantly alter their globally random coil character.

In a recent paper Fitzkee and Rose used a Monte Carlo method to randomize a number of known protein structures to construct largely native-like ensembles that convincingly displayed random coil characteristics[15]. Specifically, for a set of 33 proteins they varied backbone torsion angles at random for a select set of ~8% of the residues, and kept the remaining ~92% of the residues in their native conformation. Thus chains in their ensembles are comprised of rigid segments of native protein structure interconnected by flexible hinge residues, similar to the chains generated by our Monte Carlo procedure except for the rigid segments being α or PPII helix in the present study. The occurrence of Flory's scaling behavior both for the ensembles of Fitzkee and Rose and for the ensembles considered in this work can be understood in the spirit of the broken rod-like chain model discussed above.

**Similarity between the Chain Dimensions for The Highly Helical, Methanol-unfolded State and The Helix-poor Urea-unfolded State – a Possible Explanation of the Puzzle.**

In contrast to the robust 3/5 exponent in the power-law relationship, the pre-exponential factor $R_0$ in Eq. 3 is rather sensitive to a number of structural factors. For example, for homopolymers, $R_0$ depends on the monomer volume, intra-monomer interaction energies, and the persistence length. Consistent with this, our simulations show that $R_0$ changes as the helical content is altered (Fig. 3). In the case of $\alpha$ helices, $R_0$ drops as $P_h$ increases up to about 80%, and then rises as $P_h$ further increases up to the pure helix limit. Over the range of $P_h$ = 0 - 90%, the absolute magnitude of $R_g$ changes up to 25% relative to the unstructured chain ensemble (Fig. 4). For PPII helices, in contrast, we see a monotonic increase of $R_g$ with increasing $P_h$. The magnitude of the variation in the chain dimension caused by the presence of PPII structure is generally larger than that for $\alpha$ helices.

Whereas the dimensions of poly-alanine chains derived using the pure hard shell repulsion Ramachandran map are somewhat larger than the dimensions of experimentally characterized, unfolded proteins, the mean dimensions of the chains generated by biasing towards a slightly higher helix content ($P_h$ ~20%) coincide quite closely with the experimentally observed $R_g$ (Fig. 3A). It is interesting to note that the hard shell repulsion Ramachandran map underestimates the helical propensities of poly-alanine as compared to the Ramachandran map derived by extracting $\{\Phi,\Psi\}$s from solved proteins structures[34]. Thus it appears that poly-alanine is a good model system for the unfolded ensembles populated by the experimentally characterized proteins.

The dip of $R_g$ at an intermediate helical content is also predicted by the broken rod-like chain model (Fig. 4). The reason for such a dip is quite simple: at low $P_h$ the polymer tends to have short helical segments that are contracted compared to the coil conformations, and at high $P_h$ long helical segments tend to appear that are more expanded than the coils[35]. The $P_h$ value where the minimal chain dimension occurs depends on the geometry of the helix as well as its thermodynamic stability. The broken rod-like chain model predicts these maximally contracted conformations to occur at a much lower $P_h$ values for PPII as compared to the α helices, because a PPII helix is about two times longer than an α helix containing the same number of residues. As a consequence, the minimum of $R_g$ achieved at an intermediate value of $P_h$ appears in the MC results for residual α helices, but not in those for PPII.

In light of the above findings, an interesting question arises, *i.e.* to which extent residual local structure can be "hidden" in the scatter of experimental $R_G$ versus length plots? The capacity of a data set for distinguishing residual structural content can be estimated in the following way. Suppose that each of the collection of experimentally characterized, denatured proteins contains a unique amount of residual structure ($P_i$, $i$ = 1, 2, …, 20) so that it satisfies the 3/5 power law for its chain length $N_i$, but with a different pre-exponential factor $R_0(i)$. For simplicity further assume that $N_i$ for the 20 proteins spread uniformly over the length range $N$ = 30-600, namely $N_i$ = 30×$i$. Thus we have a series of 20 radii of gyration, $R_g(i) = R_0(i) N_i^{3/5}$, $i$ = 1, 2, …, 20. Now let us fit the 20 $R_g$ with a single power function, $R_g(i) \sim N_i^{\nu}$, and assume that the best fit yields a common pre-exponential factor $R_C$. Apparently, deviations of the individual, *actual* pre-exponential factors from the *common* one obtained from the fitting, $\Delta R_i = R_0(i) - R_C$, determine the errors of the $\nu$ values extracted from the same fit, $\Delta\nu = \nu - 3/5$. Assuming that the proteins have the same $|\Delta R_i/R_C|$ but a random sign, then the standard power law fitting procedure leads to a linear relation between the fitting error and the heterogeneity of the data set, $\Delta \nu = k \times (\Delta R_i/R_C)$, with $k = 0.305 \pm 0.002$ for $\Delta R_i/R_C \leq 30\%$. Note, however, that fits to the data set analyzed by Millet et al.[11] yield a $\Delta \nu$ of merely 0.021, which implies a data heterogeneity in the range $\Delta R_i/R_C = 5\% - 10\%$.

In our simulations, we found that, when the helical content varies over a range spanned by MeOH and urea unfolded proteins ($P_h$ = 20 - 90%), $R_g$ changes as much as 25%, implying that the current data set of denatured proteins is likely to be capable of quantitatively distinguishing residual helix abundance, at least to some extent. This makes all the more puzzling the experimental finding of a very similar chain dimension for the GuHCl/urea-induced unfolded state of low helical abundance on the one hand, and for the MeOH-unfoled, highly helical state on the other[11]. As can be seen in Fig. 3, the conformations obtained with an intermediate value of $P_h$ (*e.g.* 61%) are the most compact. Of note, there exist two ensembles, one for smaller $P_h$ (*e.g.* 21%) and the other for larger $P_h$ (*e.g.* 87%), which are characterized by almost indistinguishable $R_g$. (In fact, both ensembles agree equally well with the single power law curve that best fits the experimental data.) If the two ensembles may be assigned respectively to the GuHCl/urea-induced unfolded state and the MeOH-unfoled state, the puzzle mentioned above will be resolved naturally. Essentially, the similar dimension for both denatured states of distinctly different helix abundance might be simply the result of a coincidence emerging from the non-monotonic dependence of the persistence length on the helical content. Future improvement of the quality of the data set will provide a better chance to test this explanation.

**PPII Helix in Denatured Proteins**

It has been suggested on both theoretical and experimental grounds that PPII helices are an important contributor to the unfolded ensemble[8, 18, 19, 36]. However, experimentally it is not easy to unambiguously identify PPII structure in a denatured protein of large *N*, in which a variety of other conformational elements are usually present as well. The distinctly monotonic, and rather dramatic increase of $R_g$ with increasing PPII abundance may contribute a new type of signal for a better identification of PPII in denatured proteins in future experiments. As can be seen in Fig. 3B the chains generated by assuming PPII helix as the sole residual structure are always more expanded than the dimensions observed experimentally. These results seem to suggest that PPII is unlikely to be the dominant preferred conformation for unfolded proteins. Nevertheless, it is possible for a large amount of PPII helix to exist in denatured proteins, but only if other kinds of residual structure are also present which affect chain dimension in a compensating way. For example, α helices may coexist with PPII structure[37], balancing each other's influence on chain dimensions.

## ACKNOWLEDGEMENT

This work was supported by grants from the Robert A. Welch foundation and by the NSF CAREER award (to DEM) and by NIH grant R01GM62868-01A2 (to KWP). The work was also partly supported by National Natural Science Foundation of China (Grant No. 90403006, to ZSW).

**Figure Captions**

Figure 1. A Ramachandran map for the alanine di-peptide. The triangles indicate the ideal $\{\Phi,\Psi\}$ pairs for $\alpha$ and PPII helices, respectively.

Figure 2. Power law scaling. Panel A: logarithmic plot of $R_g$ as a function of chain length ($N$) for three representative conformational ensembles from the simulation: the unstructured chain ensemble (filled circles), an ensemble with 63% extra $\alpha$ helical content relative to the unstructured chain ensemble (filled squares), and an ensemble with 50% extra PPII content (empty squares). The lines are power law fitting. Panel B: exponents extracted from the fitting as a function of $\alpha$ helical contents (filled squares). The bars give fitting errors. The shadow area indicates the range of experimental values estimated from small angle scattering data of denatured proteins[11].

Figure 3. Comparison between simulation and experiment. The squares are experimental data adopted from the literature[11]. The lower solid line is power law fit of the data, while the upper solid line is a simulation result for the unperturbed, $P_h = 0$, ensemble. Panel A: Simulation results for ensembles of different average $\alpha$ helical contents. Panel B: results for ensembles of different average PPII contents.

Figure 4. Chain dimension as a function of residual helix contents for $N = 240$ poly-alanine. The squares are simulation results with the filled squares for $\alpha$ helices, and the open ones for PPII. The lines are predictions of Eq.1 with the solid line representing $\alpha$ helices, and the dashed line representing PPII.

**Table 1.** Minimum contact distances used in this study (in Å). Values in parentheses are used when a hydrogen bond is formed.

|     | N    | $C_\alpha$ | C′   | O              | H    | HN             |
|-----|------|------------|------|----------------|------|----------------|
| N   | 2.57 | 2.85       | 2.71 | 2.57 (2.33)    | 2.32 | 2.32           |
| $C_\alpha$ |      | 3.14       | 2.99 | 2.85           | 2.52 | 2.52           |
| C′  |      |            | 2.85 | 2.71           | 2.38 | 2.38           |
| O   |      |            |      | 2.57           | 2.32 | 2.32 (1.71)    |
| H   |      |            |      |                | 1.90 | 1.90           |
| HN  |      |            |      |                |      | 1.90           |

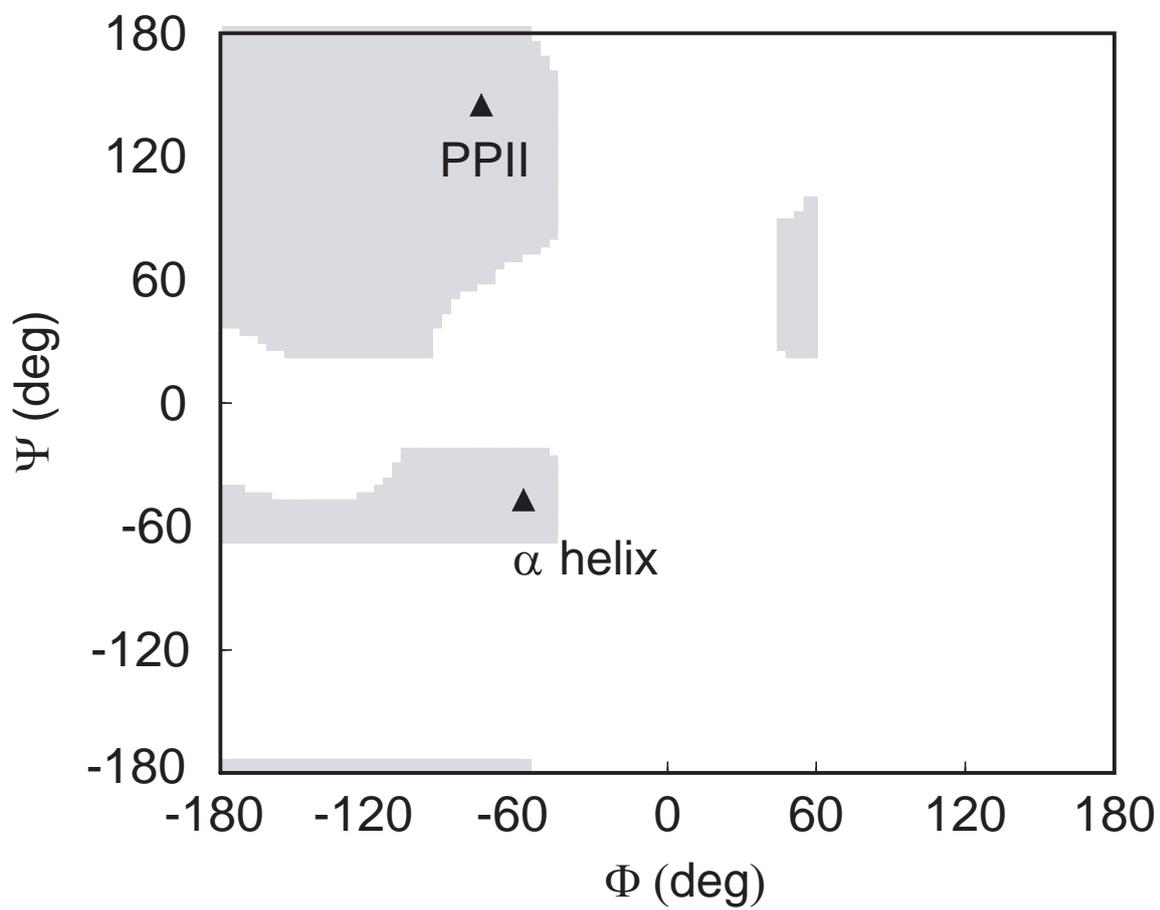

Wang, Plaxco and Makarov  Figure 1

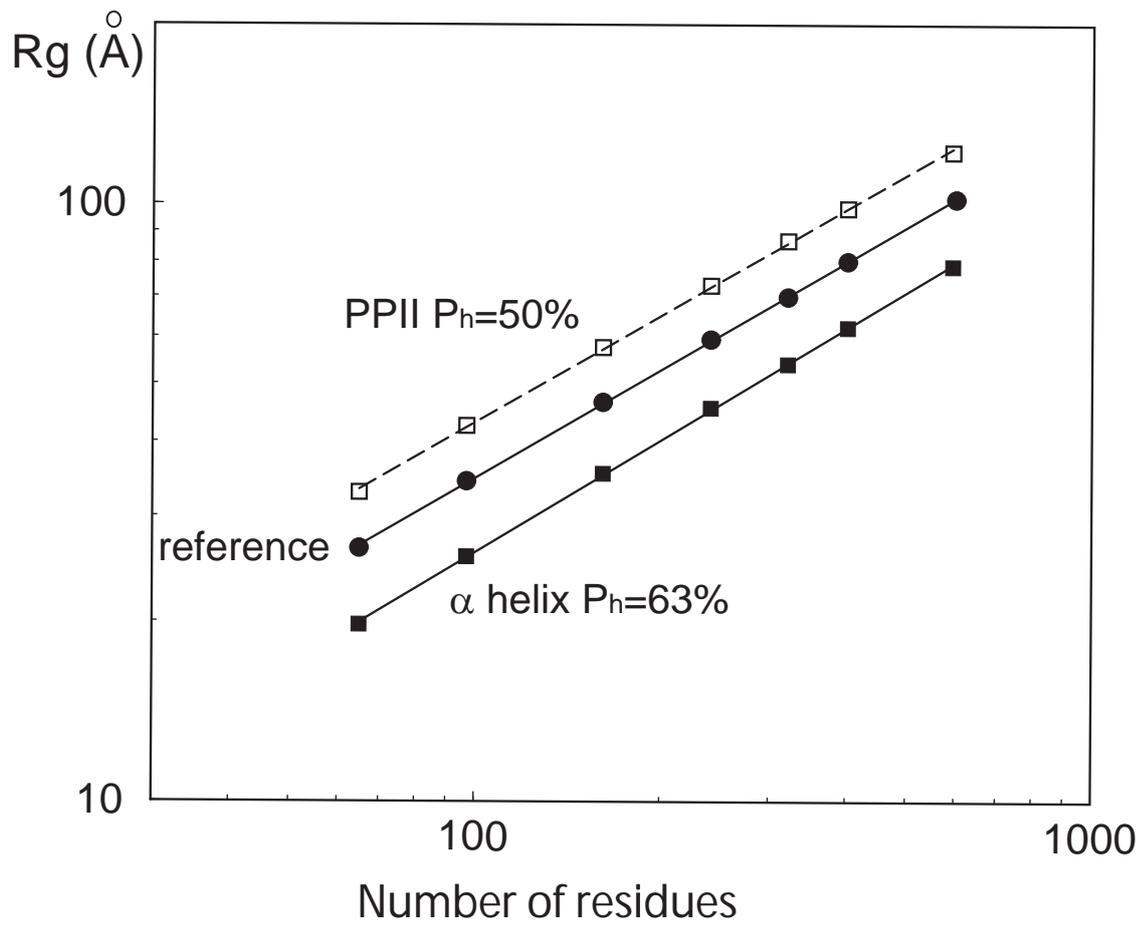

Wang, Plaxco and Makarov Figure 2A

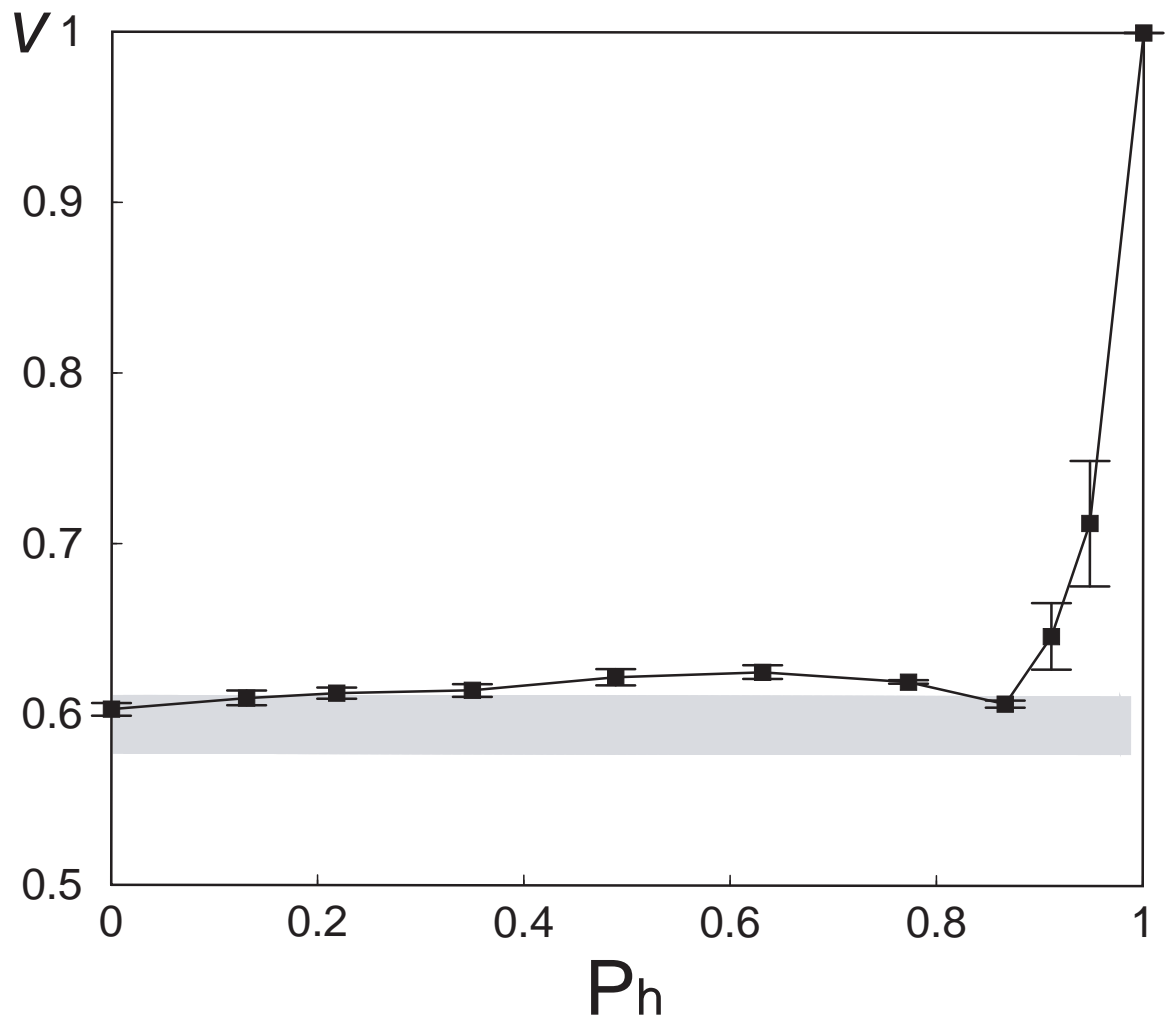

Wang, Plaxco and Makarov Figure 2B

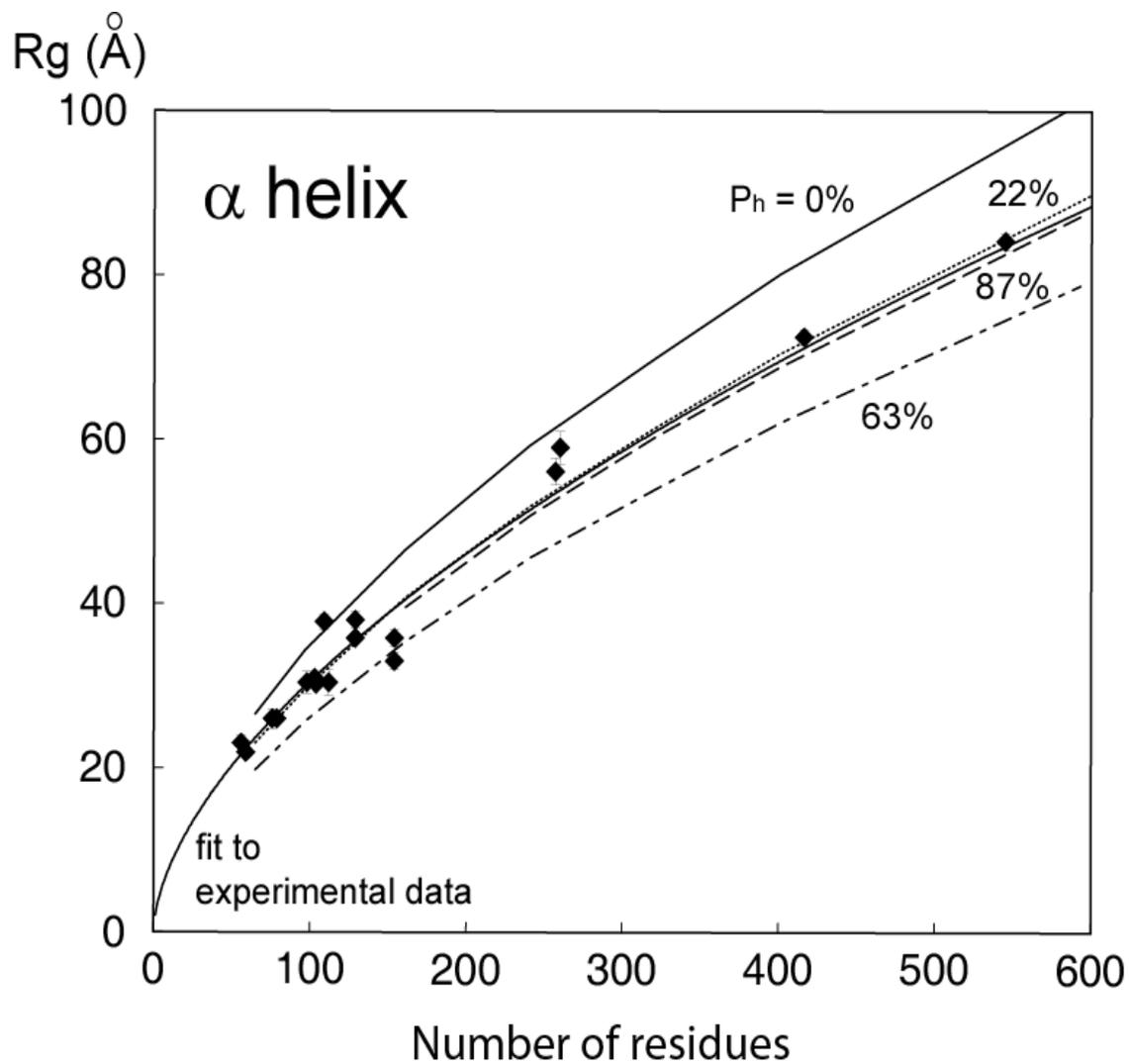

Wang, Plaxco and Makarov Figure 3A

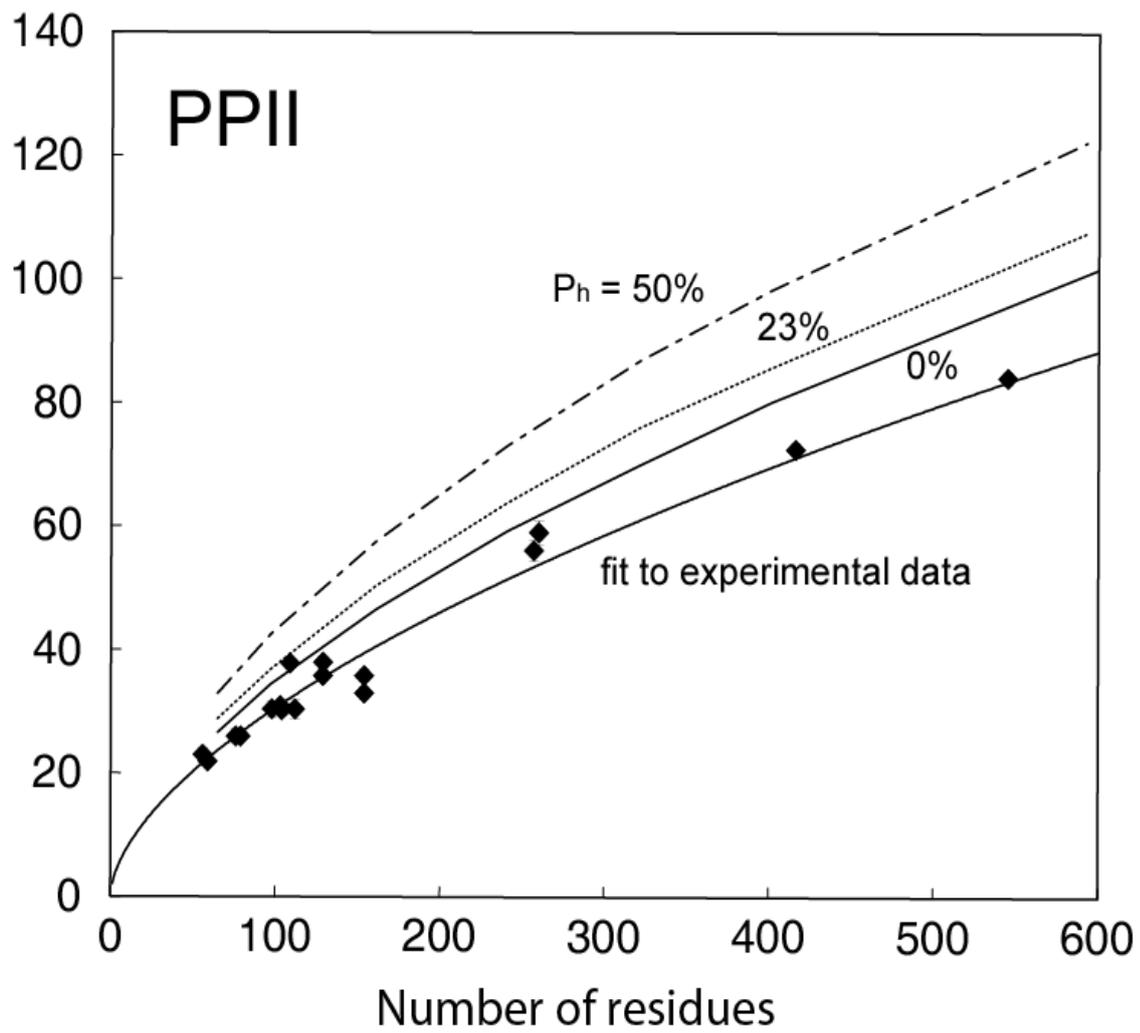

Wang, Plaxco and Makarov Figure 3B

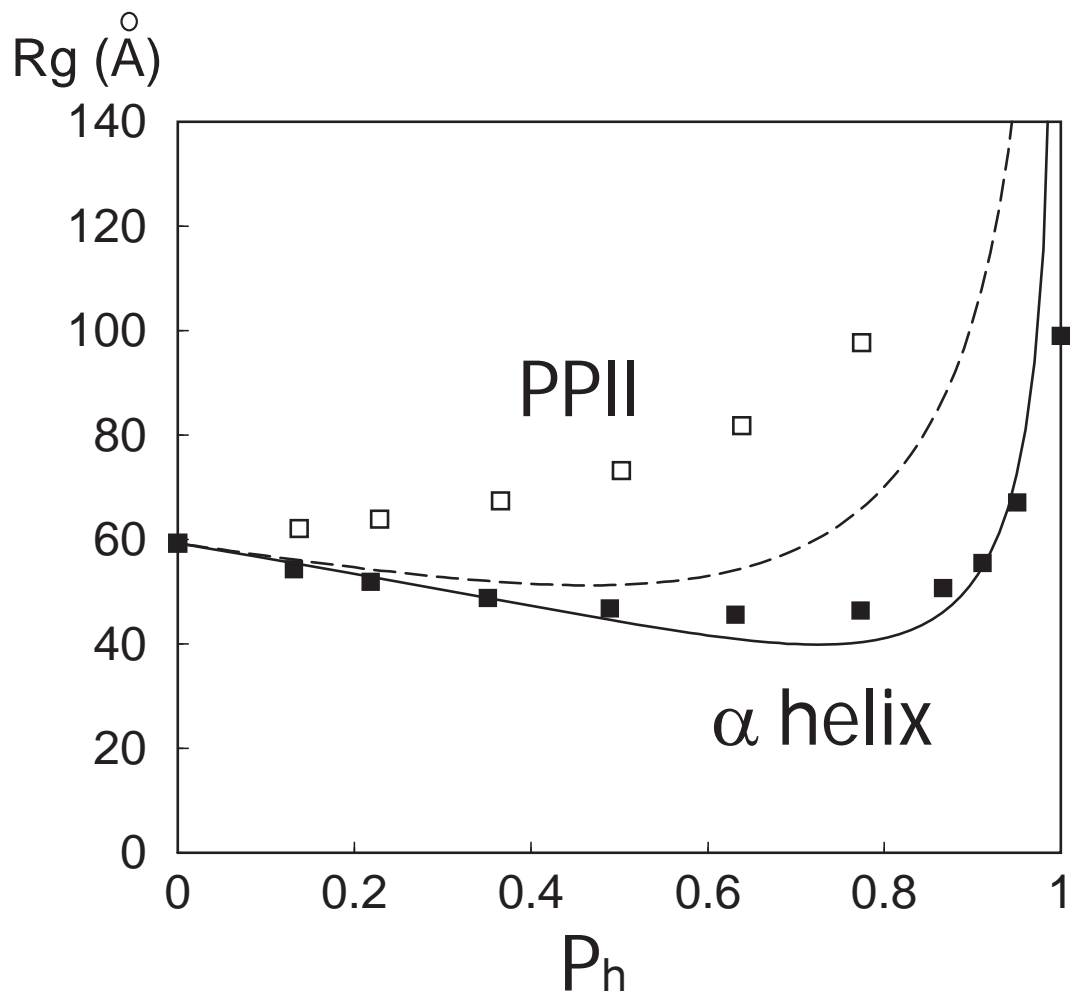

Wang, Plaxco and Makarov Figure 4